\documentclass[a4paper,11pt]{article}
\usepackage{jcappub} 
\usepackage{hyperref}
\usepackage{lineno}
\usepackage{natbib}

\newcommand{\ep}{$\epsilon$}
\newcommand{\od}{$\Omega_{\rm dust}$}

\newcommand{\sdt}[1]{{\color[rgb]{0.7,0.6,0.5} {\textbf{#1}}}}

\usepackage[dvipsnames]{xcolor}
\usepackage{xcolor}

\title{\boldmath Implications for dark energy of cosmic transparency in light of DESI data}

\author{S. Dhawan,}
\affiliation{School of Physics \& Astronomy and Institute of Gravitational Wave Astronomy\\
University of Birmingham, UK}
\author{E. M{\"o}rtsell }
\affiliation{Oskar Klein Centre, Department of Physics, Stockholm University, \\ Albanova University Center, SE-106 91 Stockholm, Sweden}

\emailAdd{s.dhawan@bham.ac.uk}

\abstract{
The distance duality relation (DDR) between luminosity and angular diameter distances holds if gravity is described by a metric theory and the universe is transparent. Recent cosmological inferences using Type Ia supernovae (SNe~Ia), baryon acoustic oscillation (BAO) and the cosmic microwave background (CMB) observations have suggested that dark energy may evolve in time. We test how the assumption of distance duality impacts dark energy inference.  Marginalizing over the absolute SNe~Ia luminosity, we find no deviation from the DDR, independent of the SN~Ia compilation used, or the assumed dark energy model. This corresponds a maximum deviation in the SN~Ia luminosity of $\Delta m \sim 0.05$ mag at the highest redshift. Allowing for deviations in the DDR increases the errors in the dark energy equation of state parameters (EoS) by 30-50$\%$. For the Pantheon+ compilation the constraints on dark energy are within $2\sigma$ of $\Lambda$CDM when applying a more realistic minimum redshift cut $z_{\rm min} >0.023$. 
We constrain possible physical scenarios that can impact cosmic transparency, specifically photon-axion mixing and the presence of (gray) intergalactic (IG) dust, in the latter case limiting the dust density to $\Omega_{\rm dust}<2 \times 10^{-4}$ at 95\% C.L. When using the local Cepheid calibration of the SNe~Ia absolute luminosity, a significant deviation from the DDR relation (for which $\epsilon=0$) is preferred. However, the best fit parameter value $\epsilon = -0.091 \pm 0.024$ requires the SNe~Ia to be brighter than the case with DDR valid, therefore, neither of the physical scenarios which dim the SNe can explain the high $H_0$.
Constraints on the photon-axion interaction length scale suggest a limit on the coupling constant of $g_{a\gamma} < 10^{-12}\,{\rm GeV}^{-1}$. 
These constraints are expected to improve  by a factor of $\sim 2$ with future data, e.g., from the Roman Space Telescope and stage-IV BAO surveys.

}

\begin{document}
\maketitle
\flushbottom

\section{Introduction}
\label{sec:intro}

The Type Ia supernova (SN~Ia) magnitude-redshift relation shows that the universe is accelerating \cite{1998AJ....116.1009Riess,1999ApJ...517..565Perlmutter}.
This current era of accelerated expansion, driven by a currently unknown dark energy component, has been a cosmic surprise. The physical origin of dark energy has profound implications for cosmology and particle physics, with several possible scenarios posited to explain the observations \cite{2025arXiv250314743Lodha,2025arXiv250204929Wolf,2025PhRvD.111d1303Wolf,2025arXiv250320178UrenaLopez,2025arXiv250319898Pan}. A combination of different probes, including the SN~Ia Hubble diagram \cite{Brout2022,Rubin2023,2024ApJ...973L..14DES}, BAO and the CMB \cite{2020A&A...641A...6Planck}, is used to infer the properties of dark energy, including its present day EoS ($w_0$) and its time-dependence ($w_a$). The orthogonality between the different probes, due to different redshift ranges and kinds of distances involved, allows for simultaneous constraints of $w_0$ and $w_a$. Recent results from the Dark Energy Spectroscopic Instrument (DESI) collaboration's second data release, DR2 \cite{desicollaboration2025desidr2resultsii}, indicate a strong ($> 4 \sigma$) hint that the dark energy density is not constant in time - unlike a cosmological constant for which $w_0=-1$ and $w_a=0$. If the result is not due to systematic uncertainties, this is a sign that dark energy is more exotic than previously assumed. However, the inference requires an assumption on how different types of distance measures relate to each other. 

The  two most common physical properties used for inferring distances are the brightness and the angular size, probing the luminosity and angular diameter distance, respectively \cite[e.g., see][]{Hogg1999}. 
The conservation of the phase space density of photons in Lorentz invariance metric garvity theories requires that the luminosity distance is related to angular diameter distance by a factor of $(1+z)^2$. As a fundamental symmetry, this relation - known as the Etherington relation, the Tolman test, or DDR  - is independent of the cosmological model, or even the assumption of homogeneity and isotropy \cite{Etherington1933}. 
While analyses testing the validity of the DDR have assumed phenomenological models for dark energy \cite{Renzi2022}, some of the dark energy parametrisations allow for unphysical dynamics of the expansion history. This can be overcome in models of dark energy dynamics governed by a physically motivated scalar field, e.g. thawing quintessence \cite{2025arXiv250415190Dinda}.  

The DDR depends on cosmic transparency being properly accounted for, e.g., interactions of photons with dust or a dark sector. 
Photons can be attenuated by, e.g. intergalactic dust or the gas and plasma in and around galaxies that absorbs, scatters and re-emits  a fraction of the incident radiation light at longer wavelengths, see e.g. \citep{2012MNRAS.426.3360J}. A more exotic scenario is if, at least part, of the dark matter (DM) is in the form of axions or axion-like particles (ALP) interacting with photons, 
vhere photon-axion conversion can take place in the presence of external magnetic fields \citep{Mirizzi2005,Mirizzi2008}. Such photon-axion oscillations can be constrained using the magnitude-redshift relation for well-calibrated distance indicators, e.g. SNe~Ia \citep{Mortsell:2002dd, Bassett2004a,Bassett2004b, Buen-Abad2020} and the differential attenuation of quasars \citep{Mortsell:2003ya,Ostman:2004eh}. 
Recent analyses of CMB polarisation data hints at ($\sim 3 \sigma$) new physics beyond the standard
model that violates parity symmetry, termed as cosmic birefringence \citep{Eskilt2022,Eskilt2023}. It is, therefore, interesting to test using later epoch data, whether there are signatures of such ``axion-like" pseudoscalar fields coupling to electromagnetism - which could manifest as a deviation from the DDR.

With current observations, there is a significant tension between the Hubble constant, $H_0$, derived from local measurements \citep{Riess2022} and the value inferred from CMB observations \citep[CMB;][]{2020A&A...641A...6Planck}. The local measurements, known as the “local distance ladder”, are based on the calibration of the absolute luminosity of SNe~Ia using independent distance measurements to host galaxies of nearby SNe~Ia. This tension, if confirmed, could provide evidence for new fundamental physics beyond the standard model of cosmology. While there are several potential explanations suggested for beyond the standard model cosmology \citep{2022PhR...984....1Schonberg, 2018JCAP...09..025Mortsell,2016JCAP...10..019Bernal}, there are no concordance explanations of the Hubble tension. Proposed solutions can be broadly categorised into two sub-classes, typically known as early or late-time solutions, based on whether they modify the physics before recombination (i.e., at redshifts $z \sim 1100$) or in the last $\sim$ 10 billion years of expansion history (i.e., $z \lesssim 2$). A common threads in a majority of the models is that they modify the expansion rate as a function of redshift, i.e., $H(z)$. Testing the validity of the distance duality relation is an independent channel to interpret the inference of dark energy parameters and  search for a cosmological solution to the Hubble tension.

 In this paper, we investigate the relation between deviations of the DDR and dark energy EoS inferences, in light of new observations of the SN~Ia magnitude-redshift relation, BAO and the local distance ladder. We assume both phenomenological and physically motivated parametrisations of the expansion history that avoid unphysical phantom crossings, where the dark energy EoS crosses $w=-1$. 
 We first test the consistency of the DDR using SNe~Ia with uncalibrated absolute luminosities together with BAO data. We then quantify the impact of the local and CMB calibrations of the distance scale on the inferred deviations from the DDR. We then analyse how physical effects could violate the DDR and constrain the photon-axion coupling and the intergalactic dust density. Finally, with stage-IV missions like the Roman Space Telescope \cite{2018ApJ...867...23Hounsell} and the Rubin Observatory \cite{thelsstdarkenergysciencecollaboration2021lsstdarkenergyscience}, soon to begin taking data, we forecast how constraints on the photon-axion coupling will improve in the near future. 

\section{Inference, Methodology and Dataset}
\label{sec:data}
In this section, we setup the inference methodology that we use for testing the impact of the DDR on dark energy inference. We define the datasets used in the analysis and the methodology for parameter inference. 

\label{eq:dr_dist}


\subsection{Type Ia supernovae}
For our analysis we use the SN~Ia magnitude-redshift relation for which there are several compilations in the literature with large samples and detailed analyses of the systematics. Here, we use the Pantheon+\cite{Brout2022}, Union \cite{Rubin2023} and Dark Energy Survey (DES; \cite{2024ApJ...973L..14DES}) compilations, to test the impact that the choice of the SN~Ia sample has on the final result. Since peculiar velocities corrections at low redshifts are very uncertain, we only use SNe~Ia for which $z > 0.023$; the same selection cut used for the Hubble flow in inferring $H_0$ \cite{Riess2022}. This selection does not impact the Union and DES samples. However, it differs from the selection employed for the Pantheon+ sample in, e.g., \cite{desicollaboration2025desidr2resultsii}.    
The distance modulus predicted by a homogeneous and isotropic, flat Friedman-Robertson-Walker (FRW) universe is given by
\begin{equation}
\mu(z; \theta) = 5\, \mathrm{log_{10}} \left( \frac{d_{\rm L}}{10\, \mathrm{Mpc}} \right) + 25
\label{eq:mu_sne}
\end{equation}
where $z$ is the redshift, $\theta$ are the cosmological parameters  \citep[e.g., see][]{Brout2022} and the luminosity distance $d_{\rm L}$ is given by 

\begin{equation}
d_{\rm L} = d_{\rm A} (1 + z)^{(2+\epsilon)}.
\label{eq:lum_dist}
\end{equation}
Here, $\epsilon$ parametrises the deviation from the Etherington distance duality relation. The impact on the SN~Ia magnitude at a given redshift is, therefore, $\Delta m = 5\epsilon {\rm log} (1+z)$. the angular distance $d_{\rm A}$ is given by 
\begin{equation}
    d_{\rm A} = \frac{c}{H_0 (1+z)}\int_0^z\frac{dz}{E(z)}\, ,
\label{eq:d_A}
\end{equation}
\begin{figure}
    \centering
    \includegraphics[width=0.48\linewidth, trim = 20 0 20 0]{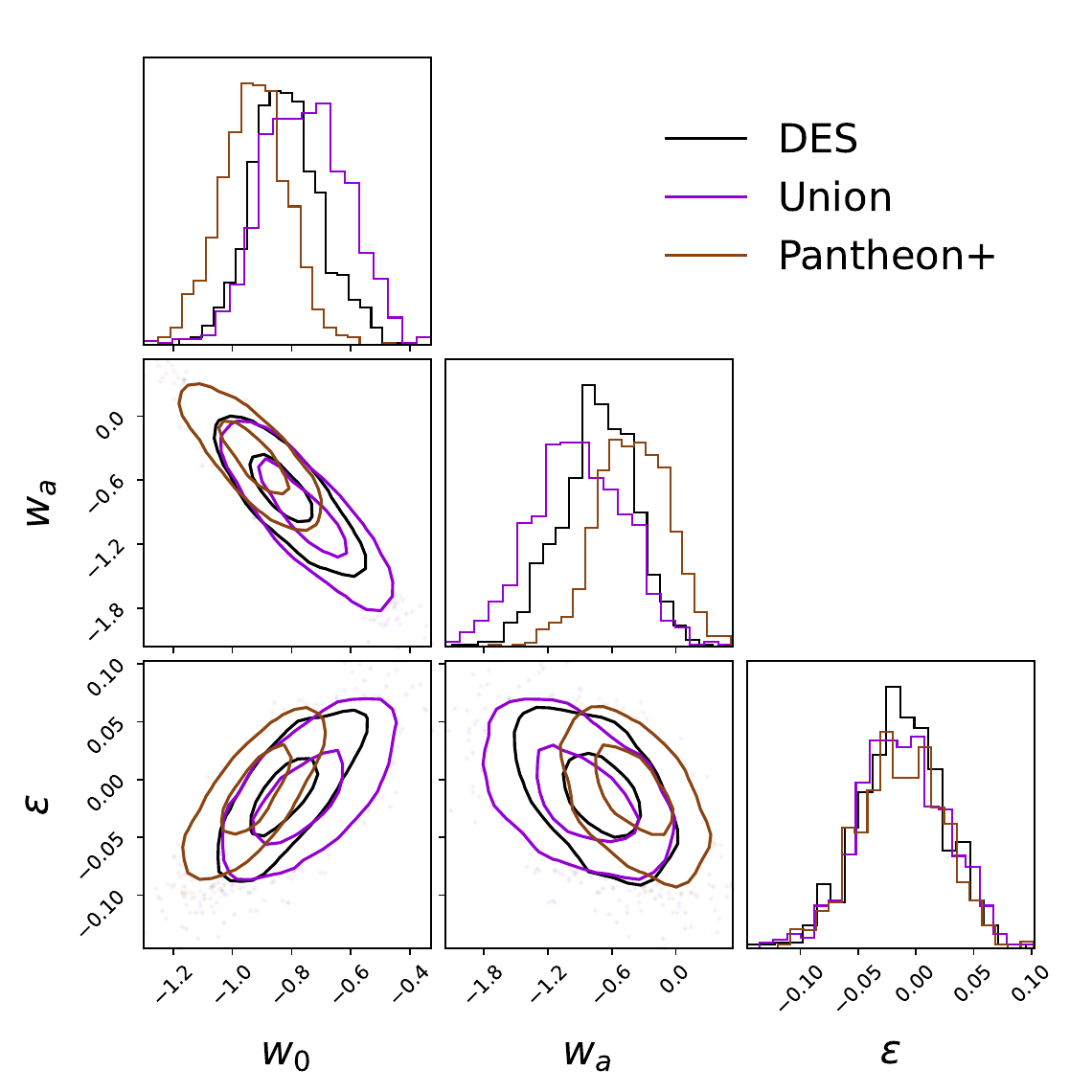}
    \includegraphics[width=.48\textwidth, trim = 20 0 20 0]{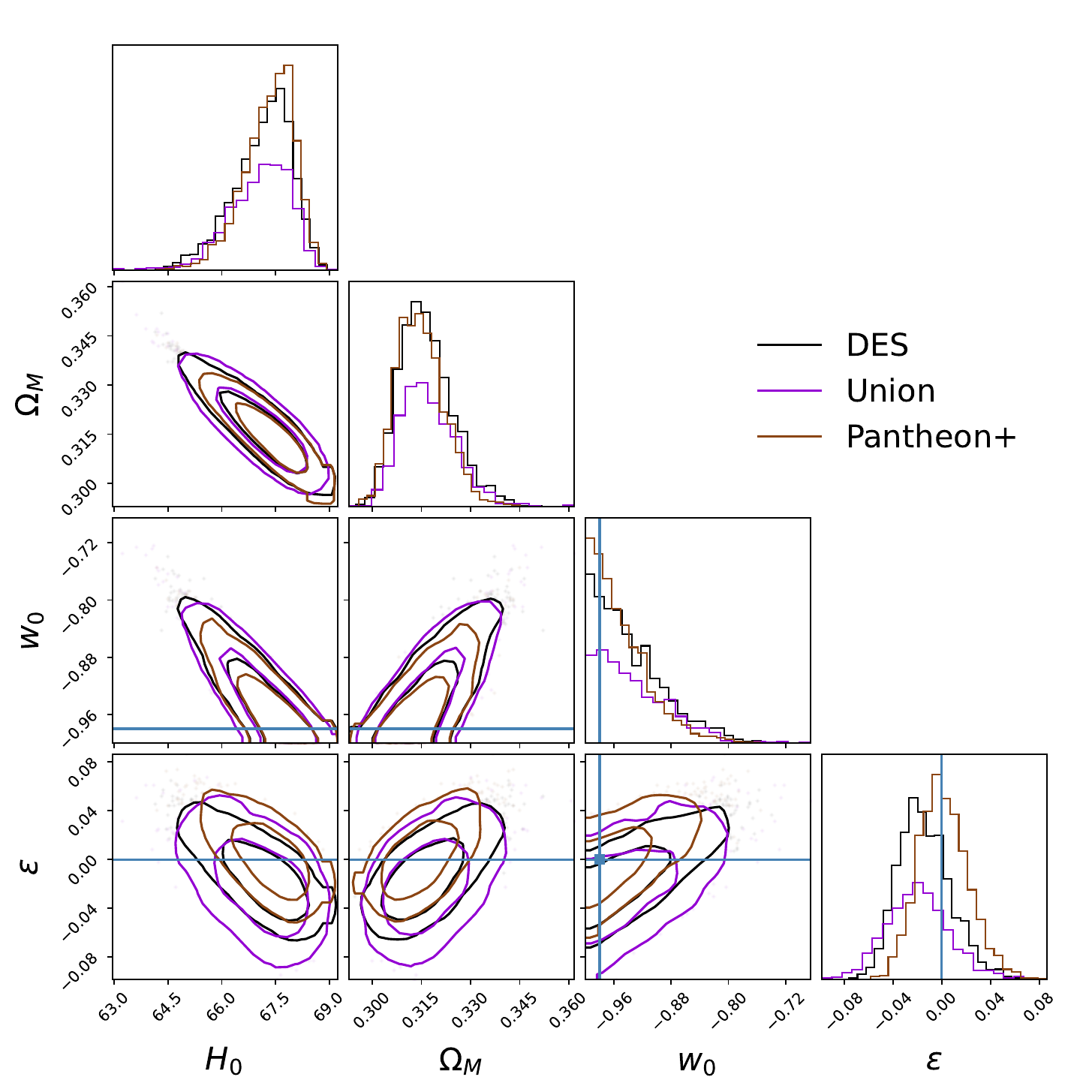}
    \caption{Constraints on the deviation from distance duality from each of the three SN~Ia samples, Pantheon+ (brown), Union (violet) and DES (black). While the Union and DES samples indicate deviations from a cosmological constant at high significance, the Pantheon+ data is consistent with $(w_0, w_a)=(-1, 0)$ at the $< 2\sigma$ level. However, independent of the SN~Ia sample, there is no deviation from distance duality (left panel). Right panel: Results for a thawing quintessence background cosmology, wherein we show $\epsilon$ along with the $\Omega_{\rm M}$ and $w_0$ constraints.}
    \label{fig:des_union_comp}
\end{figure}
where $E(z)\equiv H(z)/H_0$ is given by 
\begin{equation}
E^2(z)   = \Omega_\mathrm{M} (1+z)^3 + \Omega_{\mathrm{DE}}(z) + \Omega_\mathrm{K}(1+z)^2, 
\label{eq:norm_hubbleparameter}
\end{equation}
with
\begin{equation}
\mathrm{\Omega_{\mathrm{DE}}(z)} = \Omega_{\mathrm{DE}}\, \mathrm{exp} \left[ 3 \int_0^z \frac{1+w(x)}{1 + x} dx \right].
\label{eq:ode_z}
\end{equation}
Here, $w(z)$ is the dark energy EoS. 

Observationally, the distance modulus is calculated from the SN~Ia peak apparent magnitude ($m_B$), light curve width ($x_1$) and colour ($c$)
\begin{equation}
\mu_{\rm obs} = m_B - (M_B - \alpha x_1 + \beta c) + \Delta_\mathrm{M} + \Delta_\mathrm{B},
\label{eq:obs_distmod}
\end{equation}
where $M_B$ is the absolute magnitude of the SN~Ia. Here, $\alpha$, $\beta$, are the slopes of the width-luminosity and colour-luminosity relation. $\Delta_\mathrm{M}$ and $\Delta_\mathrm{B}$ are the host galaxy mass step and distance bias corrections respectively \citep[see, ][for details ]{Brout2022}. 

Assuming normally distributed uncertainties, we sample over the input parameters and using a $\chi^2$ given by 
\begin{equation}
\chi_{\mathrm{SN}}^2 = \Delta^T C_{\mathrm{SN}}^{-1} \Delta, 
\end{equation}
where $\Delta = \mu - \mu_{\rm obs}$ and $C_{\mathrm{SN}}$ is the covariance matrix described in \cite{Brout2022}.  We sample over the parameters and compute the likelihood using a nested sampling algorithm, MultiNest as introduced in \cite{2009MNRAS.398.1601F, 2014A&A...564A.125B}, as well as with emcee, an affine invariant Markov chain Monte Carlo (MCMC) ensemble sampler, giving equivalent results.

Due to the degeneracies in the parameter inference from the SNe~Ia distance-redshift relation, we combine the SN~Ia data with complementary probes, as described below. 

\subsection{Cosmic Microwave Background}
For the CMB we use a compressed likelihood from the \emph{Planck} satellite, with the CMB shift parameter, $R$, the angular position of the first acoustic peak in the power spectrum, $l_{\rm A}$ and the baryon density at present day, $\Omega_{\rm b} h^2$ comprising the data vector, \cite[see][ for details]{Bansal2025}.
The CMB shift and the position of the first acoustic peak are given by
\begin{equation}
R = \sqrt{\Omega_\mathrm{M} H_0^2} d_{\rm A}(z_*)/c,
\label{eq:cmb_shift}
\end{equation}
and
\begin{equation}
l_{\rm A} = \pi \frac{d_{\rm A}(z_*)}{r_s(z_*)}
\label{eq:cmb_la}
\end{equation}
where $r_s(z)$ is the sound horizon at the CMB decoupling redshift $z_*$. The data vector for the compressed likelihood is \cite{Bansal2025} 
\begin{equation}
    \nu_{\rm CMB} = [R, l_A, \Omega_b h^2] = [1.7541, 301.77, 0.022371]
\end{equation}
with corresponding covariance matrix (scaled by $10^{8}$)
\begin{equation}
C_{\rm CMB} = 
\begin{pmatrix}
  1559.83 & -1325.41 & -36.45  \\ 
    -1325.41 &  714691.80 &  269.77 \\
    -36.45 &  269.77 & 2.10     \\
\end{pmatrix}  
\end{equation}
The likelihood $\mathcal{L}_{\rm CMB}$is given by
\begin{equation}
    -0.5{\rm log} (\mathcal{L}) = \chi^2 = \Delta_{\rm CMB}^T C^{-1} \Delta_{\rm CMB}
\label{eq:likelihood_cmb}
\end{equation}
where $\Delta_{\rm CMB} = \nu_{\rm CMB, obs} - \nu_{\rm CMB. model}$. In \cite{Bansal2025}, a detailed comparison of the inference of dark energy parameters with the full CMB likelihood and the compression are performed, finding excellent agreement for probe combinations like the ones used in this work. 

\vspace{-0.2cm}
\subsection{Baryon Acoustic Oscillation}
Observation of the characteristic scale of the BAO in the correlation function of different matter distribution tracers provides a powerful standard ruler to probe the distance-redshift relation. BAO analyses constrain a few different distance measurements depending on the dataset, namely, $D_{\rm H}$, $D_{\rm M}$, and $D_{\rm V}$, i.e., the Hubble, comoving and volume distances, which we collectively refer to as $D_x$. $D_M$ is defined as $D_A (1+z)$, where $d_A$ is defined in eq.~\ref{eq:d_A}, $D_H = R_H / E(z)$ and $D_V = (z D_M^2 D_H)^{1/3}$. {
Here, we use the latest data from the Dark Energy Spectroscopic Instrument  second data release \citep[DESI;][]{desicollaboration2025desidr2resultsii} \footnote{An easy, machine-readable form of the dataset used can be found here: \hyperlink{bao}{https://github.com/sdhawan21/BAODataCovariance}} which report values of $D_V$, $D_M$ and $D_H$. 
The $\chi^2$ is given by 
\begin{equation}
    \chi^2 = (d_X - d_{X,th}){^T} C_{\rm BAO}^{-1} (d_X - d_{X,th})
\end{equation}
where $d_X = D_X / r_d$. 
Following the formalism for the sound horizon at drag redshift, $r_d$ from approximations to results from a full Boltzmann code analysis, we use the expression
\begin{equation}
    r_d = 147.05  \left(\frac{\omega_m}{0.1432}\right)^{-0.23} \left(\frac{\omega_b}{0.02236}\right)^{-0.13}  {\rm Mpc}
\label{eq:rs_approx}
\end{equation}
assuming a $N_{\rm eff} = 3.04$, following \cite{DESI-VI, 2023JCAP...04..023B}. Here, $\omega_m = \Omega_M h^2$ and $\omega_b = \Omega_b h^2$, corresponding to physical total matter and baryon densities.

\begin{table}[]
\caption{Parameters of the DDR model fit to various probe combinations and SN~Ia compilations, assuming a CPL expansion history.}
\resizebox{\textwidth}{!}{
\begin{tabular}{|c|c|c|c|c|c|c|c|}
\hline
Probe combination & $\Omega_M$ & $H_0$ & $w_0$ & $w_a$ & $\epsilon$ & $\chi^2$ & $N_{\rm dof}$\\
\hline 
AllProbes & 0.279 $\pm$ 0.007 & 71.84 $\pm$ 0.90 & -1.11 $\pm$ 0.09 & -0.15 $\pm$ 0.32 & -0.09 $\pm$ 0.025 & 1238.4 & 1457 \\
CMB-BAO-SN & 0.307 $\pm$ 0.013 & 68.50 $\pm$ 1.48 & -0.93 $\pm$ 0.11 & -0.40 $\pm$ 0.33 & -0.014 $\pm$ 0.037&1223.36 & 1380 \\
All-NoDDR & 0.311 $\pm$ 0.006 & 68.00 $\pm$ 0.61 & -0.89 $\pm$ 0.06 & -0.49 $\pm$ 0.29 & $\ldots$&1252.54 & 1381 \\
DES & 0.314 $\pm$ 0.013 & 67.68 $\pm$ 1.45 &  -0.82 $\pm$ 0.12 & -0.71 $\pm$ 0.35 & -0.012 $\pm$ 0.035 & 1649.83 & 1839 \\
Union & 0.322 $\pm$ 0.015 & 66.91 $\pm$ 1.55 & -0.75 $\pm$ 0.14 & -0.91 $\pm$ 0.42 & -0.008 $\pm$ 0.038 & 33.46 & 31 \\
\hline 
\end{tabular}}

\end{table}

\section{Results}
\label{sec:results}
Here, we present constraints on the dark energy model parameters, simultaneously inferred with the DDR parameter $\epsilon$. We infer the parameters in the time-dependent dark energy parametrisation given by the Chevallier-Polarski-Linder (CPL) \cite{2001IJMPD..10..213Chevallier, 2003PhRvL..90i1301Linder} formalism wherein $w(a) = w_0 + (1-a) w_a$  , as well as the simplified $\Lambda$CDM model to test how much the expansion history model impacts our constraints. We then test the role of the local distance ladder calibration on the inferred cosmology. 

\subsection{CPL parametrisation of dark energy} 
With current results on $w_0-w_a$ suggesting deviations from $(-1, 0)$, it is important to test the assumptions going into the inference framework. Constraining $\epsilon$ in a CPL parametrisation for dark energy with all three SN~Ia compilations, we find $\epsilon = -0.014 \pm 0.037$, $-0.012 \pm 0.035$, $-0.008 \pm 0.038$ for the Pantheon+, DES and Union compilations, respectively. A summary of the constraints is presented in Figure~\ref{fig:des_union_comp}. 

 Within this model we find ($w_0= -0.972 \pm 0.114$, $w_a= -0.395 \pm 0.331$), for Pantheon+, ($w_0=-0.748 \pm 0.141$, $w_a=-0.91 \pm 0.422$) for Union and ($w_0=-0.823 \pm 0.116$, , $w_a=-0.713 \pm 0.345$) for DES. We note that the combination of using a higher $z_{\rm min}$ compared to \cite{desicollaboration2025desidr2resultsii} and allowing for DDR violations in the for of $\epsilon\neq 0$, the Pantheon+ constraints are consistent with $\Lambda$, i.e., $(w_0, w_a)=(-1,0)$, at the $< 2 \sigma$ level.  
\begin{table}[]
\caption{Parameters of the DDR model fit to various probe combinations, assuming a thawing quintessence expansion history from \cite{2024MNRAS.533.2615Camilleri}.}
\resizebox{\textwidth}{!}{\begin{tabular}{|c|c|c|c|c|c|c|}
\hline
Probe Combination & $\Omega_M$ & $H_0$ & $w_0$ & $\epsilon$ & $\chi^2$ & $N_{\rm dof}$\\
\hline 
AllProbes & 0.301 $\pm$ 0.005 & 68.58 $\pm$ 0.37 & -0.99 $\pm$ 0.01 & -0.033 $\pm$ 0.016&1259.07 & 1458 \\
CMB-BAO-SN & 0.315 $\pm$ 0.007 & 67.34 $\pm$ 0.73 & -0.96 $\pm$ 0.04 & 0.002 $\pm$ 0.022&1225.26 & 1381 \\
All-NoDDR & 0.313 $\pm$ 0.005 & 67.53 $\pm$ 0.50 & -0.96 $\pm$ 0.03 & $\ldots$&1225.62 & 1382 \\
DES & 0.316 $\pm$ 0.008 & 67.25 $\pm$ 0.82 & -0.95 $\pm$ 0.05 & -0.016 $\pm$ 0.021 &1654.92 & 1840 \\
Union & 0.316 $\pm$ 0.008 & 67.16 $\pm$ 0.81 & -0.94 $\pm$ 0.05 & -0.020 $\pm$ 0.025 &38.74 & 32 \\

\hline 
\end{tabular}}
\label{tab:thaw}

\end{table}
\subsection{Thawing Quintessence}

While the phenomenological CPL parametrisation encapsulates a wide range of physical models, it also includes several unphysical regions of parameter space, e.g., ones with negative kinetic energy or ``ghost" modes arising at phantom crossings \cite{2017PhRvD..96f3524Peirone,2024PhRvD.110l3516Mukh, 2025arXiv250415190Dinda}. Recent studies have also shown that the inference of the time dependence of dark energy is sensitive to reparametrisations of $w(z)$ \cite{2025arXiv250502658Efstathiou}. Therefore, it is imperative to test more physically motivated models of dark energy. Here, we test a thawing quintessence model, proposed as a possible explanation of deviations from the $\Lambda$CDM model \cite{2025arXiv250204929Wolf}. The nomenclature for thawing/freezing models was introduced in \cite{2005PhRvL..95n1301Caldwell} - where thawing refers to the EoS of dark energy being frozen to $w=-1$ in the early universe and then ``thaws" to less negative values. Forecasts have shown that these models can be constrained very well by complementary probes in the context of next generation missions, e.g., time-delay distances from strong gravitational lensing \cite{2025arXiv250206929Shajib}. While different dynamical dark energy potentials, $V(\phi)$, have slightly different behaviours of the EoS with redshift  \cite{2017JCAP...07..040Dhawan}, e.g., algebraic thawing \cite{2015PhRvD..91f3006Linder} and pseudo-nambu-Goldstone-Bosons (PNGB; \cite{1995PhRvL..75.2077Frieman,2017JCAP...01..023SmerBarreto}) models can be summarised by an EoS \cite{2024MNRAS.533.2615Camilleri}
\begin{equation}
    w(z) = -1 + (1 + w_0) e^{-\alpha z}
\label{eq:eos_thaw}
\end{equation}
where $\alpha = 1.45 \pm 0.1$.
Constraints on the dark energy parameters and distance duality are presented in Table~\ref{tab:thaw} and Figure~\ref{fig:des_union_comp} (right panel).

\subsection{Local Calibration of SN~Ia magnitudes}
Since the Cepheid calibrated local distance ladder measurement of $H_0$ is in tension with the inference from the CMB, we do not by default combine the calibrated SN~Ia magnitude data with CMB observations. Here, we infer the DDR parameter along with the cosmology including the local distance scale. This is to test whether a violation of the DDR could be a solution to the Hubble tension, in which case we would not be combining inconsistent datasets. In the previous sections, we used all three SN~Ia compilations for assessing the parameter inference. Here, we use only the Pantheon+ compilation, because it is the only SN~Ia compilation for which there is ans estimata of the covariance between the local and high-$z$ and SN~Ia samples.

For this dataset, $\epsilon = -0.091 \pm 0.024$, displaying a strong deviation from the DDR, 
with a $\Delta \chi^2 = 12$ in favour of the best fit DDR violation, compared to $\epsilon=0$. We find $H_0 = 71.80 \pm 0.9$ km\,s$^{-1}$\,Mpc$^{-1}$, which is closer to the local value than the early universe calibration. 
The best fit $\epsilon$ corresponds to a difference of $\sim 0.04$ mag at $z = 0.1$ and of $\sim 0.15$ mag at $z = 1$. While some studies suggest a possible deviation due to unknown systematics \cite{2025arXiv250416868Afroz}, the effect of this DDR violation at high-$z$ is very large compared to the typical amplitude of systematic errors $\sim$ 0.05 mag \citep{2025MNRAS.tmp..764Dhawan}. We discuss in section~\ref{sec:discussion} whether there are unaccounted for physical phenomena that can explain this effect. A similar result has also been seen using DESI DR1 BAO data, and a gaussian prior for the local calibration in \cite{2025arXiv250410464Teixeira}.



\subsection{Validity of the Distance Duality Relation}
We have tested the DDR within the context of two different dark energy models, with three SN~Ia compilations, as well as with including the local calibration of SN~Ia absolute luminosities. For a CPL cosmology, all uncalibrated SN~Ia datasets, in combination with the CMB and BAO indicate consistency with the DDR, i.e. $\epsilon = 0$ at the $< 1 \sigma$ level. For the thawing quintessence model, we similarly find no strong deviations from the $\epsilon = 0$ case at $> 1 \sigma$ level. For the Union and DES compilations, allowing for DDR deviations yield more negative $w_0$  - i.e. closer to $-1$ by $\sim 0.02$. However, the shift is within the error margin for $w_0$.
When using the local calibration of the SN~Ia absolute luminosity, we find a strong DDR deviation  $\epsilon = -0.091 \pm 0.024$ when assuming a CPL cosmology. However, for a thawing model, the deviation from $\epsilon = 0$ is smaller, with
$\epsilon = -0.033 \pm 0.015$  This is because the more negative $\epsilon$ imply more negative $w_0$, which is naturally truncated to $w_0 = -1$  due to the physical nature of thawing quintessence.

\section{Discussion}
\label{sec:discussion}
We have tested the DDR for different dark energy models. Marginalising over the absolute SN~Ia luminosity, We find no deviations from the DDR, independent of SN~Ia sample or dark energy model assumed.  With local distance ladder calibrated SN~Ia data, there is a strong violation of the DDR assuming the CPL dark energy parametrisation. This effect is weaker when assuming dark energy is described by a thawing scalar field. Here, we discuss possible physical scenarios that impact the validity of the DDR and the constraints we expect from current and near future data.
 

 
\begin{figure}
    \centering
    \includegraphics[width=.8\textwidth]{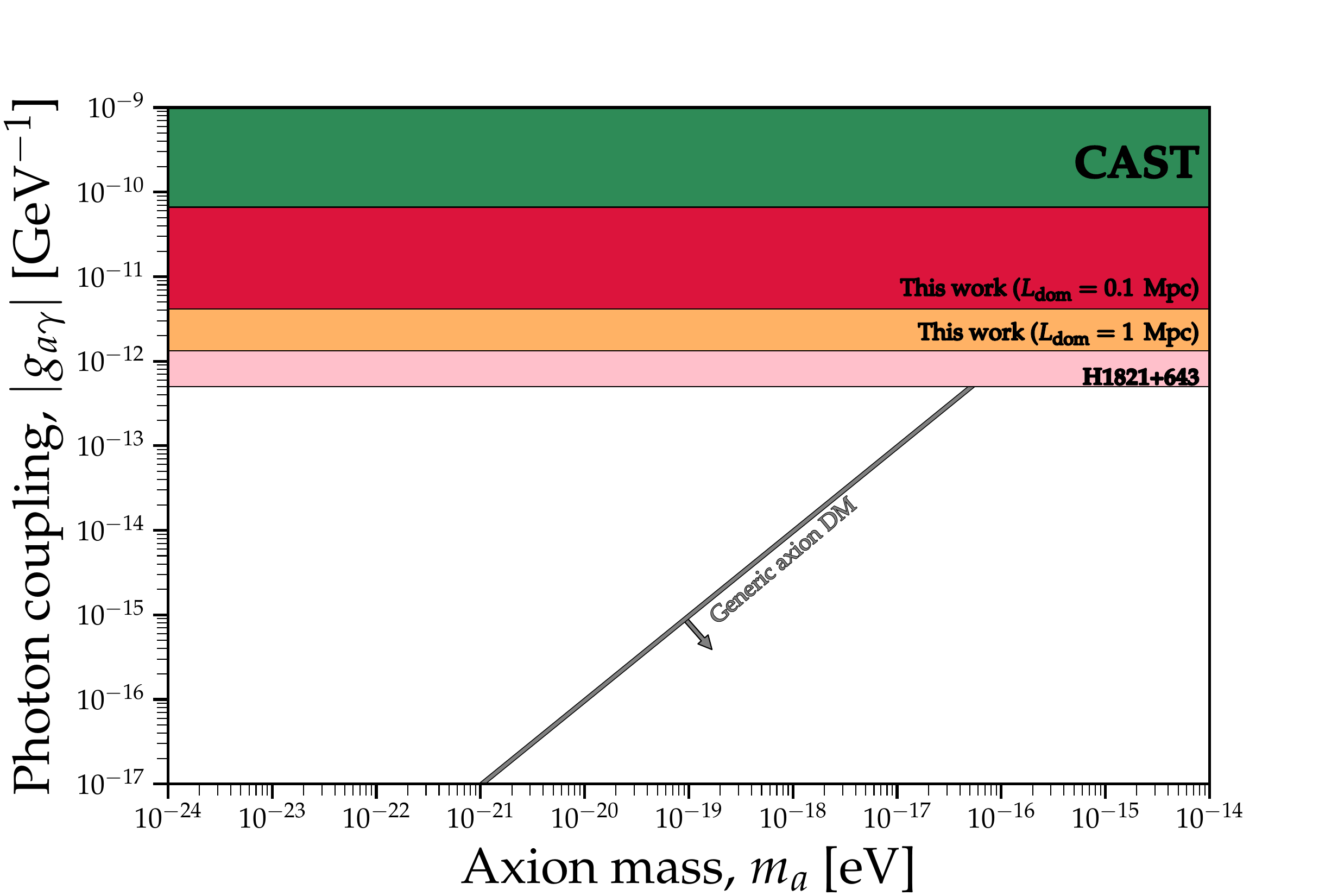}
    \caption{Comparison of the excluded region for the axion mass and the coupling between photons and axions from this work and terrestrial haloscope, e.g. CAST \cite{CAST:2017uph} and astrophysical experiments \citep{2022MNRAS.510.1264SiskReynes}. We plot the constraints with a long (mustard) and short (red) domain length to show the sensitivity of the constraints to input assumptions. The gray line shows the mass - coupling constant parameter space for which the axion can be a dark matter particle candidate. }
    \label{fig:photon-axion}
\end{figure}

\subsection{Physical effects violating distance duality}
A violation of the DDR could be due to one of a few causes, any and all of which would require modifying the propagation of photons. 
In this section we analyse some of possible physical effects that can lead to violations of the DDR. In this paper, we study photon-axion oscillations and the potential presence of intergalactic dust.

\paragraph{Photon-Axion Coupling}
It is theorised that photons may interact with particles not in
the Standard Model, e.g., axion-like particles \cite[ALPs;][]{Mortsell2002, Mukherjee2019, Buen-Abad2020} which are pseudo-scalar bosons. 
In the presence of external magnetic fields, e.g., in the intergalactic medium, photons couple to ALPs, according to the Lagrangian
\begin{equation}
    \mathcal{L_{\rm int}} = \frac{1}{4} g_{a \gamma} a F_{\mu \nu} \tilde{F}_{\mu \nu} = g_{a \gamma} a \,\,{\rm E.B}
\label{eq:lagrangian_alp}
\end{equation}
where $a$ is the ALP field, $g$ is the coupling constant and E and B are the electric and magnetic fields. In astrophysical (and cosmological) magnetic fields, photons can oscillate into ALPs and vice versa leading to changes in the photon flux, manifesting as distance-related anomalies (details in Appendix~\ref{sec:alp}).
Assuming magnetic domains of $\sim$ Mpc size with uncorrelated field direction,   the oscillation is maximal for optical photons and largely independent of energy \cite{Csaki2002a}. In the limit of infinite travel distance, one approaches an equilibrium between the two photon polarization states and the axion. The probability $P_{\gamma \gamma}$ for an initial photon being detected as a photon is given by 
\begin{equation}
P_{\gamma \gamma} = \frac{2}{3} + \frac{\exp (-l/L_a)}{3}
\label{eq:prob_gamma}
\end{equation}
where $l$ is the comoving distance to the source and $L_a$ is the exponential decay length. 
$P_{\gamma \gamma} = 1 - P_{a \gamma}$ where $P_{a \gamma}$ is the probability of conversion of a photon to an axion, as detailed in \cite{Sikivie1984,Csaki2002a}. 
We only consider the conversion in the intergalactic medium, since the conversion probability in the host galaxy magnetic field typically is $\lesssim 10^{-5}$. 
The conversion probability impacts the luminosity distance by
\begin{equation}
    D_L^{\rm eff} = D_L / \sqrt{P_{\gamma \gamma}},
\label{eq:dl_axion}
\end{equation}
causing a dimming of the observed SNe~Ia.
SN~Ia dimming is sensitive to ultra-light axions in the mass range $m_a < 10^{-10}$ eV \cite{Csaki2002a,Grossman:2002by}.

Defining the inverse of the dimensionless interaction lengthscale as $l_a^{-1} = R_H/L_a$, from the combined CMB+BAO+uncalibrated SN~Ia data we find $l_a^{-1} < 1.01$ for the CPL model and $l_a^{-1} < 0.37$ for the thawing model. This is the first attempt to constraint $l_a^{-1}$ in time-varying dark energy models. Even with the dark energy parameters free we find constraints that are a factor of 3 improved compared to previous constraints \cite{Bassett2004a}. We do not use employ the local distance ladder SN~Ia calibration since it requires SNe~Ia to be brighter and the mixing process considered here can only dim SNe~Ia. Therefore, photon loss from conversion to axions would only `` artificially tighten" the limit on $l_a^{-1}$. Hence, to be conservative we only report the limit from CMB+BAO+SNe without the local calibration. 

Assuming intergalactic magnetic field strengths of order 1 nG and domain sizes 
$L_ {\rm dom} = 1$ Mpc (as defined in section~\ref{sec:alp}), this corresponds to a limit on the coupling constant of {\bf $g_{\rm a \gamma} < 2 - 6 \times 10^{-12}\, {\rm GeV^{-1}}$} depending on the IGM magnetic field strength. The best constraints from cluster member galaxies are $g_{\rm a\gamma} < 6 \times 10^{-13} {\rm GeV}^{-1}$ \cite{2022MNRAS.510.1264SiskReynes}. The constraints from current SN~Ia samples are stronger than helioscopes. They are slightly weaker than constraints from the cluster members however they have completely different assumptions and also simultaneously marginalise over dark energy constraints.  
\vspace{-0.1cm}

\paragraph{Intergalactic Dust}
Extinction by dust in the intergalactic medium (IGM) 
would attenuate the brightness of SNe~Ia, leading to overestimates of the luminosity distances, if not properly corrected for.
However, it would leave the comoving and angular diameter distances from the BAO and CMB unchanged, thus giving rise to apparent DDR violations.
Additionally, since the effect of dimming is degenerate with the properties of dark energy, we constrain the IGM dust dust density $\Omega_{\rm dust}$ simultaneously with dark energy parameters for both CPL and thawing quintessence background cosmologies. 

In the CPL model, we find $\Omega_{\rm dust} < 2.0 \times 10^{-4}$ at the 95 $\%$ C.L., whereas the constraints are less strict for the thawing model where $\Omega_{\rm dust} < 8.9 \times 10^{-5}$. The dust density is strongly correlated with the dark energy EoS. We emphasize that these constraints are only valid for the gray component of the intergalactic dust since any colour dependent term may be over- or under-corrected for in the colour term in \ref{eq:obs_distmod}. These constraints are consistent with complementary inferences from Mg II absorbers \cite{2012ApJ...754..116Menard} and soft X-ray background observations
\cite{2009MNRAS.397.1976Dijkstra}. SNe~Ia provide complementary advantages for constraining dust owing to a sensitivity to large grain - i.e. gray - dust and observations out to high-$z$.   

\subsection{Forecasts with future Type Ia supernova surveys}
Already with current data, our constraints on the intergalactic dust density and the axion length scale are competitive and physically interesting. In the near term, we expect that observatories like the Roman Space Telescope \citep{2018ApJ...867...23Hounsell} will assemble a sample of SNe~Ia pushed to larger redshifts than current compilations. Moreover, the long wavelength baseline and improved calibration will reduce the systematic uncertainty budget \citep[see details in][]{2018ApJ...867...23Hounsell}.  
Several mission strategies are tested in \citep{2018ApJ...867...23Hounsell}, which include optimising the total time between the wide field camera for imaging and the spectrograph for spectroscopy. We take the fiducial case data products provided, i.e., the binned redshift distribution and systematic and statistical covariance matrix for our analysis. 
For the simulation input cosmology we assume a $\Lambda$CDM model with $\Omega_{\rm M} = 0.3$ and infer $\Omega_{\rm dust}$ and $l_{\rm a}$  for the dark energy models described above. 
We find $ \Omega_{\rm dust} < 1 \times 10^{-4}$ 
and  $R_H/L_a < 0.15$  
for the CPL parameterisation the 95\% C.L. limit. For the thawing model we find $ \Omega_{\rm dust} < 5.9 \times 10^{-5}$ and $l_a < 0.2$ at 95 $\%$ C.L. The constraints on \od\, will improve upon the current best limits by a factor of $\sim 2$. Since the global dust budget from \cite{2011arXiv1103.4191Fukugita} is $\Omega_{\rm dust} \sim 10^{-5}$, if there is no detection of intergalactic dust, this will have interesting implications for the method used for computing the total dust energy density budget.
The improvement in $l_{\rm a}$ by a factor of 3 is important for refining constraints on the photon-axion coupling in the mass range $m_a \lesssim 10^{-16}$eV. 
We note that while in this analysis we focussed on distance level inference of IG dust and photon-axion mixing, for future analysis, e.g., with Roman or the Vera C. Rubin Observatory \cite{thelsstdarkenergysciencecollaboration2021lsstdarkenergyscience}, these terms should be accounted for at earlier stages in the analysis. 

\section{Conclusion}
We analysed the relation between constraints on the distance duality relation and different dark energy models. For a combined dataset of BAO, CMB and different SN~Ia surveys, we find that the dark energy equation of state parameters are strongly correlated with the DDR deviation parameter, but do not find any
strong evidence for deviations from the DDR. For physically motivated dark energy models, we find slightly weaker evidence for deviations from $w=-1$ when allowing for DDR deviations. 


For the combination of Pantheon+ SNe~Ia with DESI DR2 BAO and the CMB, there is $< 2 \sigma$ evidence for departure from $w=-1$ in either CPL or thawing dark energy models. When including the local distance ladder calibration of the SN~Ia absolute luminosities, we find that in the CPL model, there is a strong ($4.5\sigma$) preference for a violation of the DDR, with a significant improvement in the goodness of fit compared to the case where the DDR is assumed to hold. In a thawing quintessence model, this departure from the DDR is weaker, at the $\sim 2 \sigma$ level.  This is because the model does not include the phantom region for the EoS of dark energy, needed for the high-$z$ probes to match the high $H_0$ to inferred from the local distance ladder calibration. 

We analysed physically motivated violations of the DDR, e.g. photon-axion oscillations and found a limit on the coupling constant of $g_{a\gamma} < 2 - 6 \times 10^{-12}$ GeV$^{-1}$, which is competitive with other astrophysical limits in the same mass range, while having complementary assumptions. 

\acknowledgments

We thank James Matthews and Pranjal Trivedi for interesting discussions on ALPs. SD acknowledges support from  UK Research and Innovation (UKRI) under the UK government’s Horizon Europe funding Guarantee EP/Z000475/1  and a Junior Research Fellowship at Lucy Cavendish College. EM acknowledges support from the
Swedish Research Council under Dnr VR 2020-03384.


\bibliographystyle{JHEP}
\bibliography{main.bib}
\appendix 
\section{Deriving Photon-axion coupling}
\label{sec:alp}
In this section, we present the formalism used to infer the photon-axion coupling from the inferred axion lengthscale.
Photon-axion oscillations can cause dimming of distant SNe~Ia which can be written in magnitudes by taking the logarithm of equation~\ref{eq:dl_axion}.
\begin{equation}
    \delta m_{\rm a} = -2.5 {\rm log}_{10}  \left[ \frac{2}{3} + \frac{\exp(-l/L_a)}{3} \right].
\label{eq:dust_rhl}
\end{equation}
\begin{figure*}
\includegraphics[width=.48\textwidth]{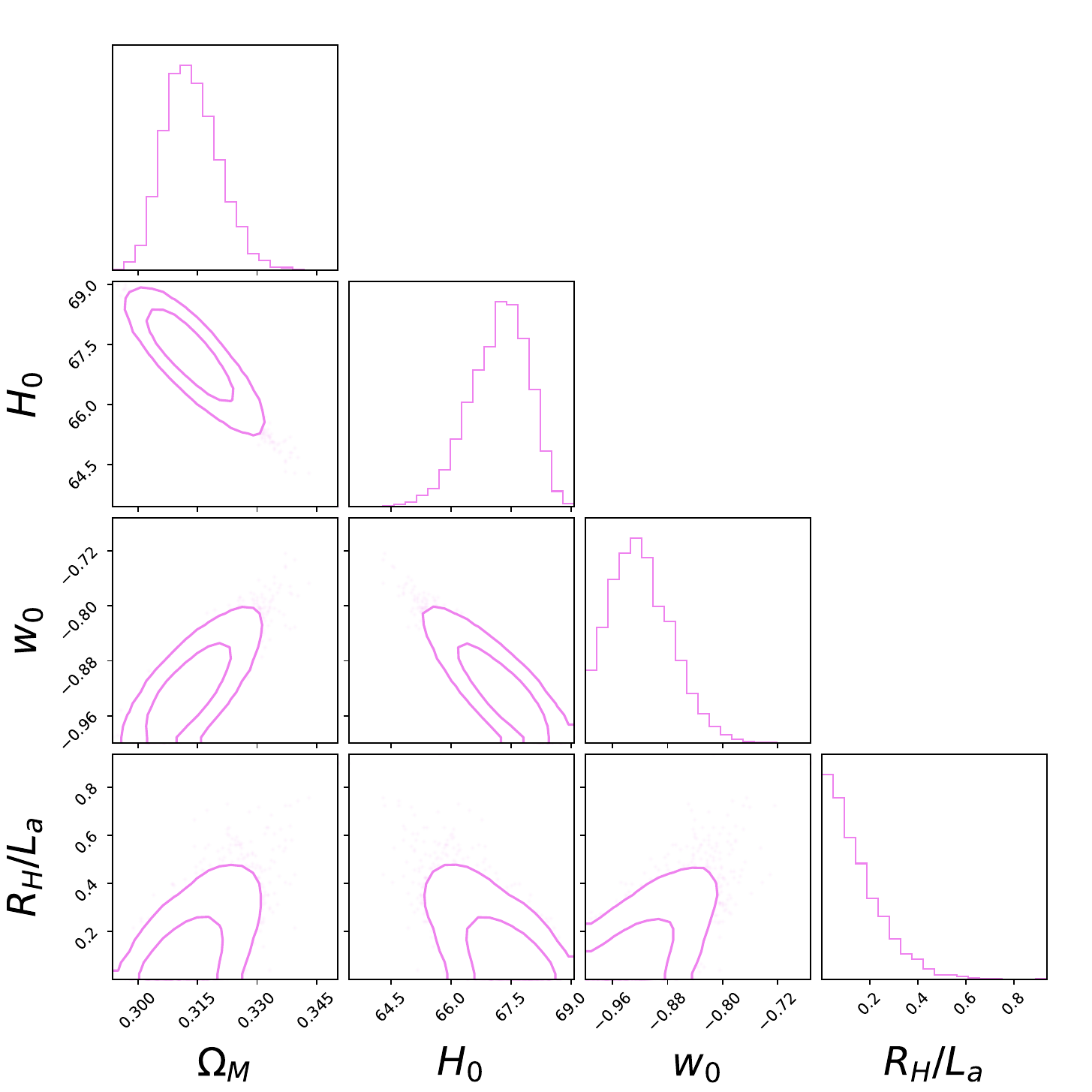}
\includegraphics[width=.48\textwidth]{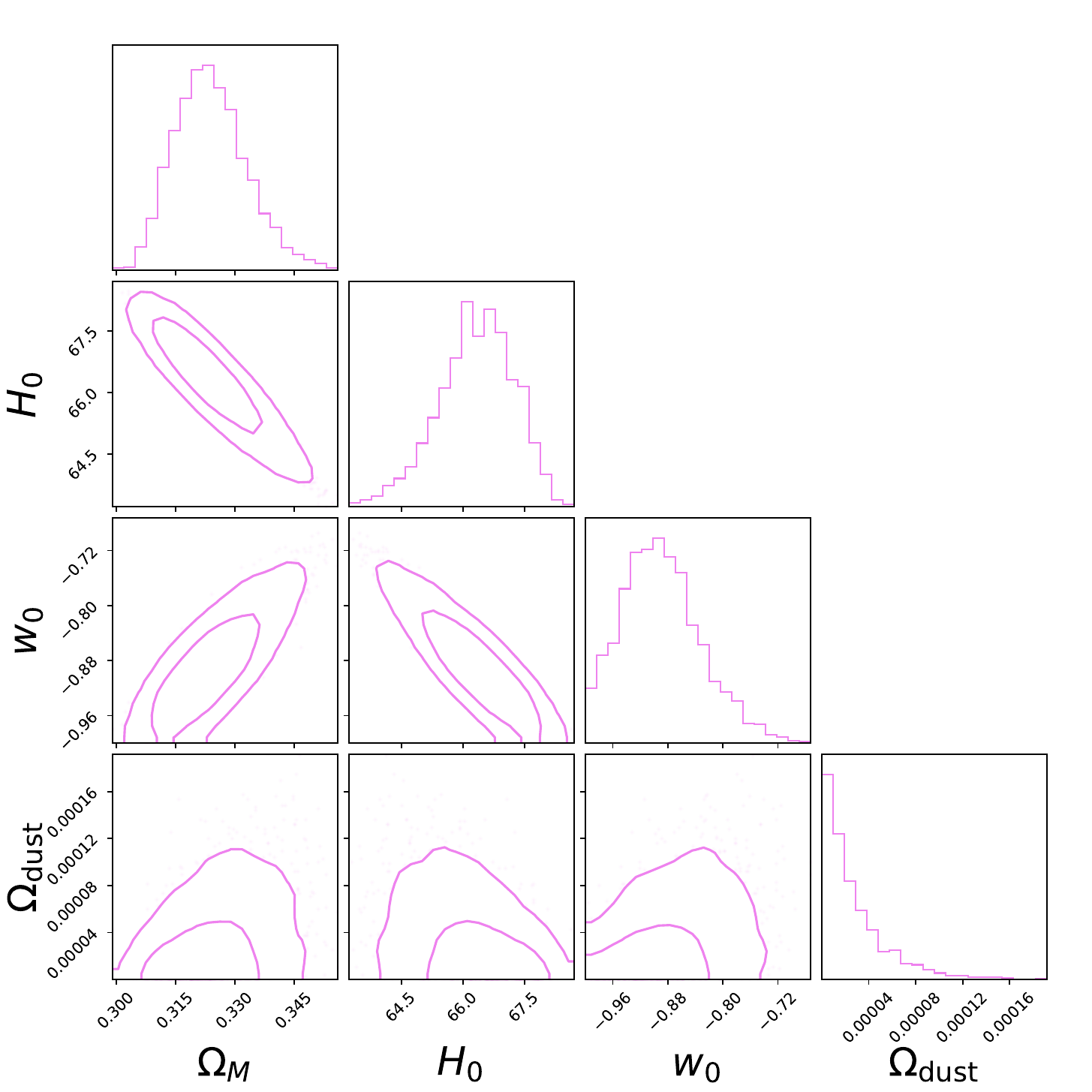}
\caption{(Left): Constraints using a combination of CMB+BAO+SNe on the photon-axion interaction lengthscale as defined in section~\ref{sec:alp}. (Right): Constraints on the amount of intergalactic dust ($\Omega_{\rm dust}$). Both physical scenarios assume a thawing quintessence model. }
\label{fig:appendix_thaw}
\end{figure*} 
From these constraints on the lengthscale $L_a$, we can infer limits on the axion mass ($m_a$) and photon-axion coupling constant $g_{\rm a, \gamma}$ by assuming an intergalactic magnetic field strength $B$ and domain size $L_{\rm dom}$. We note that for the CMB to remain unaffected by the oscillations, the mass $m_a \sim 10^{-16}$ eV or lower. The decay length $L_{\rm a}$ \cite{Csaki2002a,Grossman:2002by} corresponds to 
\begin{equation}
    L_{\rm a} = \frac{8}{3 \mu^2 L_{\rm dom}}
\end{equation}
where 
\begin{equation}
    \mu = 3.09\cdot 10^{-2}\left(\frac{B}{10^{-9}\,{\rm G}}\right)\left(\frac{M_{\rm a}}{10^{11}\,{\rm GeV}}\right)^{-1}\,{\rm Mpc}^{-1}
\end{equation}
and
\begin{equation}
L_{\rm a} = 2793 \left(\frac{B}{10^{-9}\,{\rm G}}\right)^{-2}\left(\frac{M_{\rm a}}{10^{11}\,{\rm GeV}}\right)^2\left(\frac{L_{\rm dom}}{1\,{\rm Mpc}}\right)^{-1}\,{\rm Mpc}    
\end{equation}
where $B$ is the strength of the intergalactic magnetic field and $L_{\rm dom}$ is the domain size. We assume a field strength of 1 nG and a domain size of 1 Mpc as done in previous studies \cite{Buen-Abad2020}. 


\section{IGM Dust Attenuation}
\label{sec:dust}
Here, we define the formalism for the intergalactic dust extinction used to constrain the dust density $\Omega_{\rm dust}$. 
The absorption unaccounted for in the $B$-band is given by 
\begin{equation}
    A_\lambda = \frac{2.5}{\ln 10} \frac{3 c H_0}{8\pi G} \Omega_{\rm dust} \mathcal{I}
\end{equation}
where $\mathcal{I}$
\begin{equation}
    \mathcal{I} = \int_0^{z_s} \frac{\kappa  (1+z)^{\gamma - 1}}{E(z)}.
\end{equation}
Here, $\kappa$ is the wavelength dependent mass absorption coefficient (or opacity), reflecting the nature of the dust size
and composition, for which we adopt the Small Magellanic Cloud value of $\kappa =1.54 \times 10^3$ cm$^2$g$^{-1}$ from \cite{Weingartner2001}, similar to \cite{Goobar2018}.   This is because if we assume non-gray dust - i.e. a smaller $R_V$ - the SALT2 colour correction could have over or under corrected the impact of IGM dust. $\gamma$, which determines the redshift evolution of the dust density is a free parameter in our fits.  $E(z)$ (eq.~\ref{eq:norm_hubbleparameter}) depends on the dark energy model assumed which in this study is either the CPL or thawing quintessence model. 

\end{document}